\documentclass[twocolumn,showpacs,preprintnumbers,amsmath,amssymb]{revtex4}
\usepackage{graphicx}
\begin{document}

\title{Half-integer Shapiro steps at the 0\,--\,$\pi$ crossover of a ferromagnetic Josephson junction}

\author{Hermann Sellier$^{1,2}$}
\author{Claire Baraduc$^1$}
\author{Fran\c{c}ois Lefloch$^1$}
\author{Roberto Calemczuk$^1$}
\affiliation{$^1$D\'epartement de Recherche Fondamentale sur la
Mati\`ere Condens\'ee, CEA-Grenoble, 17 rue des Martyrs, 38054
Grenoble, France\\
$^2$Kavli Institute of Nanoscience, Delft University of
Technology, Lorentzweg 1, 2628 CJ Delft, The Netherlands}

\date{accepted for publication in Phys Rev Lett, received 10 February 2004}

\begin{abstract}

We investigate the current-phase relation of S/F/S junctions near
the crossover between the 0 and the $\pi$ ground states. We use
Nb/CuNi/Nb junctions where this crossover is driven both by
thickness and temperature. For a certain thickness a non-zero
minimum of critical current is observed at the crossover
temperature. We analyze this residual supercurrent by applying a
high frequency excitation and observe the formation of
half-integer Shapiro steps. We attribute these fractional steps to
a doubling of the Josephson frequency due to a $\sin(2\phi)$
current-phase relation. This phase dependence is explained by the
splitting of the energy levels in the ferromagnetic exchange
field.

\end{abstract}


\maketitle


The current-phase relation of ballistic and diffusive S/N/S
junctions is predicted to be strongly non-sinusoidal at zero
temperature in contrast to that of tunnel junctions. This is due
to a different conductivity mechanism involving Andreev bound
states created in the normal metal (N) by the superconductors (S).
These states are sensitive to the superconducting phase difference
$\phi$ and carry the supercurrent $I_S$. In S/F/S junctions the
current-phase relation is strongly distorted by the exchange field
of the ferromagnet (F) and can even be {\it reversed} leading to
the famous $\pi$ state \cite{buzdin-82-jl}. The microscopic
mechanism responsible for this {\it negative} supercurrent can be
intuitively explained in the clean limit where the Andreev
spectrum is discrete \cite{dobrosavljevic-00-pc, golubov-02-jl,
sellier-03-prb}. The two spin configurations of each bound state
are indeed split by the ferromagnetic exchange energy. When the
first bound state is shifted from finite energy to zero energy (at
$\phi=0$), the direction of the total supercurrent given by the
lowest level (for $\phi>0$) is {\it negative} (instead of
positive). In this case the ground state is at $\phi=\pi$.

\begin{figure}[bp]
\includegraphics[width=7cm,clip,trim=0 0 0 0]{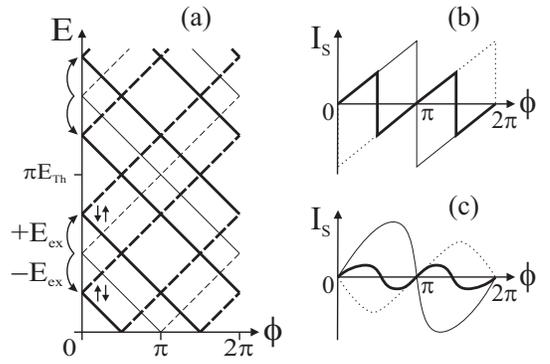}
\caption{(a) Energy of Andreev levels {\it versus} phase
difference in a one-dimensional ballistic junction ($E_{Th}=\hbar
v_F/d \ll\Delta$) \cite{sellier-03-prb}. Each state of the normal
case (thin lines) is split by the exchange energy $E_{ex}$ of the
ferromagnet (thick lines). The states in solid (dashed) lines
carry positive (negative) currents. (b) Current-phase relation for
this model at zero temperature in the normal case (thin line), in
$\pi$ junctions (dotted line) and at the crossover (thick line).
(c) Current-phase relation in diffusive regime for the same
situations.} \label{fig:phi}
\end{figure}

In this article we consider the situation where the exchange
energy $E_{ex}$ is half that of the $\pi$ junction described above
for the same thickness $d$. In this case the Andreev spectrum of a
ballistic junction contains equidistant states twice closer than
usual (Fig.~\ref{fig:phi}a). As a result the supercurrent is $\pi$
periodic in phase with a saw-tooth shape at zero temperature
(Fig.~\ref{fig:phi}b) and the ground states $\phi=0$ and
$\phi=\pi$ are degenerate \cite{radovic-01-prb}. The current-phase
relation becomes more rounded in the diffusive regime
(Fig.~\ref{fig:phi}c) where the discrete spectrum is replaced by a
continuous density of Andreev states \cite{heikkila-00-el}. At
this 0\,--\,$\pi$ crossover the current-phase relation contains a
dominant $\sin(2\phi)$ component and the critical current presents
a non-zero minimum with respect to thickness or exchange energy
variations. Experimentally the critical current of S/F/S junctions
at the crossover is however so small that it was always assumed to
vanish completely. In Nb/CuNi/Nb junctions
\cite{ryazanov-01-prl,sellier-03-prb} this behavior could be
related to the strong decoherence of the magnetic alloy and in
Nb/PdNi/Al$_2$O$_3$/Nb junctions \cite{kontos-02-prl} to the
presence of the tunnel barrier.

In this Letter we report the first observation of a small non-zero
critical current at the 0\,--\,$\pi$ crossover of a Nb/CuNi/Nb
junction and show that the corresponding current-phase relation
has the expected $\sin(2\phi)$ dependence. An evidence of such a
relation could be obtained with a two junctions superconducting
loop: if one of them has a $\sin(2\phi)$ relation, the
interference pattern under magnetic field should have maxima both
at integral and half-integral flux quanta \cite{radovic-01-prb},
whereas the maxima occur only at the integral (half-integral)
values if this junction is fully in the 0 ($\pi$) state
\cite{ryazanov-01-prb,guichard-03-prl}. However we chose to
analyze the current-phase relation by studying the dynamic
behavior of a single junction. Up to now only the equilibrium
properties of the S/F/S junctions have been theoretically and
experimentally investigated. In this Letter we present the first
study of the finite voltage behavior under high frequency
excitation to reveal the harmonics of the supercurrent. In the
junction with a non-zero critical current at the crossover we
observed half-integer Shapiro steps attributed to a $\sin(2\phi)$
current-phase relation. This phase dependence reveals the level
splitting induced by the ferromagnetic exchange field.

\begin{figure}[bp]
\includegraphics[width=7cm,clip,trim=0 0 0 0]{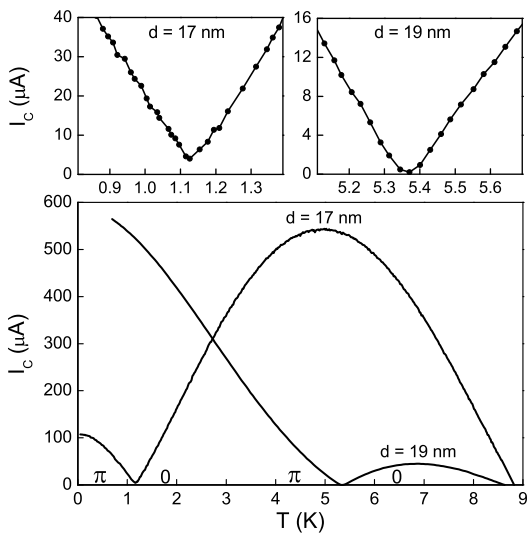}
\caption{Temperature dependence of the critical current for two
Nb/Cu$_{52}$Ni$_{48}$/Nb junctions. Upper graphs are expanded
views near the crossovers between 0 and $\pi$ states: the critical
current has a non-zero (zero) minimum for the 17 nm (19 nm) thick
junction.} \label{fig:ict}
\end{figure}

Our junctions are Nb/Cu$_{52}$Ni$_{48}$/Nb trilayers deposited
{\it in situ} and patterned by photolithography. The copper-nickel
alloy has a Curie temperature as small as 20 K. Details of these
junctions and of the experimental set-up have been reported
previously (Ref.~\cite{sellier-03-prb}). In this article we
analyze the behavior of two junctions with copper-nickel
thicknesses equal to 17 and 19 nm (the normal resistances $R_N$
are equal to 0.12 and 0.13 m$\Omega$ respectively). The
temperature dependence of their critical current $I_C$ is shown in
Fig.~\ref{fig:ict}. These unusual behaviors are explained by the
spectral supercurrent density of diffusive S/F/S junctions
\cite{sellier-03-prb,heikkila-00-el}. The ground state switches
from $\phi=\pi$ to $\phi=0$ and the critical current presents a
deep minimum at the crossover temperature $T^*$. In our set of
samples \cite{sellier-03-prb}, the 17, 18, and 19 nm thick
junctions only have this 0\,--\,$\pi$ crossover, at $T^*$
respectively equal to 1.12, 4.53, and 5.36 K. Junctions thinner
and thicker have respectively a 0 and a $\pi$ ground state at all
temperatures.

An expanded view of the crossover region reveals a finite critical
current of 4 $\mu$A at $T^*$ for the 17 nm thick junction. We show
in the following that this supercurrent has a $\sin(2\phi)$
current-phase relation. With a 0\,--\,$\pi$ crossover close to
zero temperature, this junction is indeed very close to the
optimum situation where the ratio $E_{ex}/E_{Th}$ gives a doubled
periodicity of Andreev levels. The vanishing critical current
obtained for the 19 nm (and 18 nm) thick junction may be explained
by the larger thickness giving a more stable $\pi$ state at zero
temperature. It could also be related to the higher value of
$T^*$, since the $\sin(2\phi)$ component is expected to become
much smaller when the temperature is high (of the order of the
critical temperature $T_c$ of the electrodes)
\cite{radovic-01-prb}.

\begin{figure}[bp]
\includegraphics[width=7cm,clip,trim=0 0 0 0]{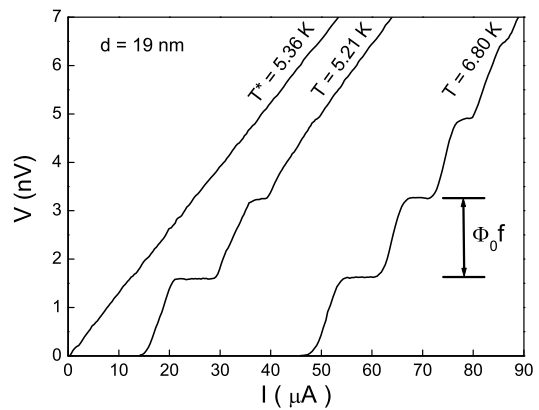}
\caption{Integer Shapiro steps in the voltage-current curve of a
19 nm thick junction with an excitation at 800 kHz (amplitude
about 18 $\mu$A). The steps disappear at the 0\,--\,$\pi$
crossover temperature $T^*$. Curves at 5.21 and 6.80 K are shifted
by 10 and 20 $\mu$A for clarity.} \label{fig:vi5kt}
\end{figure}

The current-phase relation is analyzed by sending an alternative
current at 800 kHz when measuring the direct voltage-current curve
(we use a rather low frequency because the characteristic voltage
$R_NI_C$ is in the nanovolt range as a consequence of the very
small critical current). Constant voltage Shapiro steps appear due
to the synchronization of the Josephson oscillations on the
applied excitation. The curves obtained for the 19 nm thick
junction are shown in Fig.~\ref{fig:vi5kt}. As expected the steps
appear at voltages equal to integral multiples of $\Phi_0f=1.6$ nV
($\Phi_0=h/2e$). They are similar above and below the crossover
temperature at 5.36 K where the critical current and the steps
disappear simultaneously.

\begin{figure}[bp]
\includegraphics[width=7cm,clip,trim=0 0 0 0]{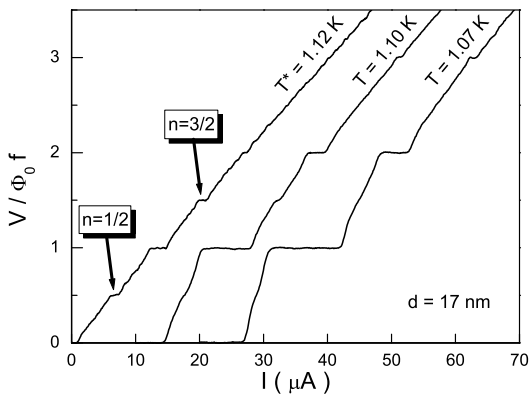}
\caption{Shapiro steps in the voltage-current curve of a 17 nm
thick junction with an excitation at 800 kHz (amplitude about 18
$\mu$A). Half-integer steps (n=1/2 and n=3/2) appear at the
0\,--\,$\pi$ crossover temperature $T^*$. Curves at 1.10 and 1.07
K are shifted by 10 and 20 $\mu$A for clarity.} \label{fig:vi1kt}
\end{figure}

The curves obtained for the 17 nm thick junction with the same
excitation amplitude and frequency are shown in
Fig.~\ref{fig:vi1kt}. The integer Shapiro steps are present at all
temperatures, including 1.12 K where the critical current is
minimum but non-zero. The new result is that two additional steps
appear at voltages $(1/2)\Phi_0f$ and $(3/2)\Phi_0f$ at this
crossover temperature. Small features at the same voltages are
also present at 1.10 K, but almost nothing is visible at 1.07 K.
These half-integer steps reveal the existence of supercurrent
oscillations at frequency $2(V/\Phi_0)$ which synchronize to the
excitation at frequency $f$ producing steps at voltages multiples
of $(1/2)\Phi_0f$. This doubling of the Josephson frequency is the
consequence of the $\sin(2\phi)$ current-phase relation expected
at the 0\,--\,$\pi$ crossover of S/F/S junctions.

Ideally, the $\sin(2\phi)$ component should dominate for this
thickness of junction at all temperatures $T\ll T_c$. The fact
that these half-integer steps are only visible close to $T^*$
indicates that the phase dependence departs very slightly from
$\sin(\phi)$. The $\sin(2\phi)$ component dominates only when the
large $\sin(\phi)$ component cancels to change its sign (i.e. when
the temperature induces the compensation of the negative and
positive currents at $\phi=\pi/2$). To the first order near $T^*$,
the supercurrent can be qualitatively described by the relation:
$I_S(T)=\frac{T-T^*}{T^*}I_1\sin(\phi)+I_2\sin(2\phi)$. The fact
that $I_1\gg I_2$ may be related to the strong decoherence in the
magnetic alloy (also responsible for the huge reduction of the
critical currents when compared with theoretical predictions
\cite{sellier-03-prb}). If the superconducting correlations are
small enough, the equations can be linearized and give indeed a
$\sin(\phi)$ relation regardless of the exchange energy. A
theoretical study of diffusive S/F/S junctions taking into account
the spin-flip scattering would be required to analyze this
behavior in more detail.

In this paragraph we present other kinds of junctions where
fractional steps have been observed in the past in order to show
the specific origin of the half-integer steps observed in our
junctions. In a current biased junction the supercurrent
oscillations at finite voltage are strongly non-sinusoidal and the
harmonics could synchronize on the excitation leading to
fractional steps. However simulations have shown that this
mechanism does not occur in the Resistively Shunted Junction (RSJ)
model with a $\sin(\phi)$ current-phase relation
\cite{hamilton-72-pla}. Fractional steps can appear if a large
capacitance or inductance is present, but this is not the case in
our non-hysteretic junctions because of their large conductance
and small critical current. Fractional steps can also appear in
case of a non-sinusoidal relation which is in fact expected from
theory in any kind of weak link at sufficiently low temperature
\cite{bardeen-72-prb,kulik-75-jl,wilhelm-98-prl}. They were
observed in superconducting micro-bridges and point-contacts, but
are more difficult to observe in diffusive S/N/S junctions: for
junctions with thick N layers compared to the coherence length,
the low temperature regime is hard to reach; and for shorter
junctions with a sandwich structure, the large cross-sections
imply large critical currents that can not be measured in the low
temperature regime (this is however possible in our case thanks to
the strong decoherence by spin-flip scattering). Fractional steps
have been observed at high temperature in long and wide diffusive
S/N/S junctions in sandwich geometries and were attributed to a
synchronization of the vortex flow \cite{clarke-68-prl}. This
flux-flow can not happen in our junctions since the Josephson
penetration length $\lambda_J$ is much larger than the junction
width. Fractional steps have also been observed in planar S/N/S
junctions at high temperature where the critical current has
vanished \cite{lehnert-99-prl,dubos-01-prl}. These steps may
involve dynamical and non-equilibrium effects acting on the
phase-coherent contribution to the resistance.

In contrast to these experiments, our half-integer steps are only
visible at the crossover, at low temperature, and with a finite
critical current. The interpretation of these half-integer steps
as a consequence of a current-phase relation with a $\sin(2\phi)$
dependence seems therefore reasonable. Since the transparency
deduced from the normal state resistance is good
\cite{sellier-03-prb}, this phase dependence has to be related to
an Andreev conduction mechanism (as opposed to tunnelling) and is
explained by the doubled periodicity of Andreev levels at the
0\,--\,$\pi$ crossover.

\begin{figure}[tbp]
\includegraphics[width=7cm,clip,trim=0 0 0 0]{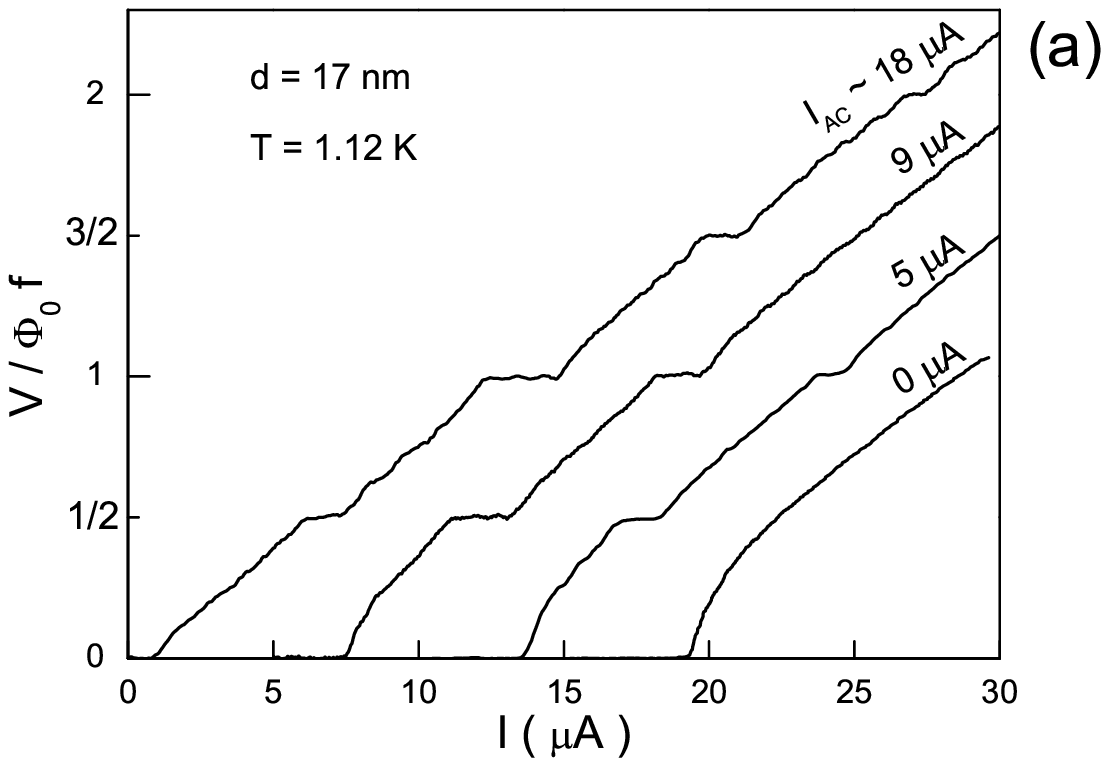}
\includegraphics[width=8cm,clip,trim=0 0 0 0]{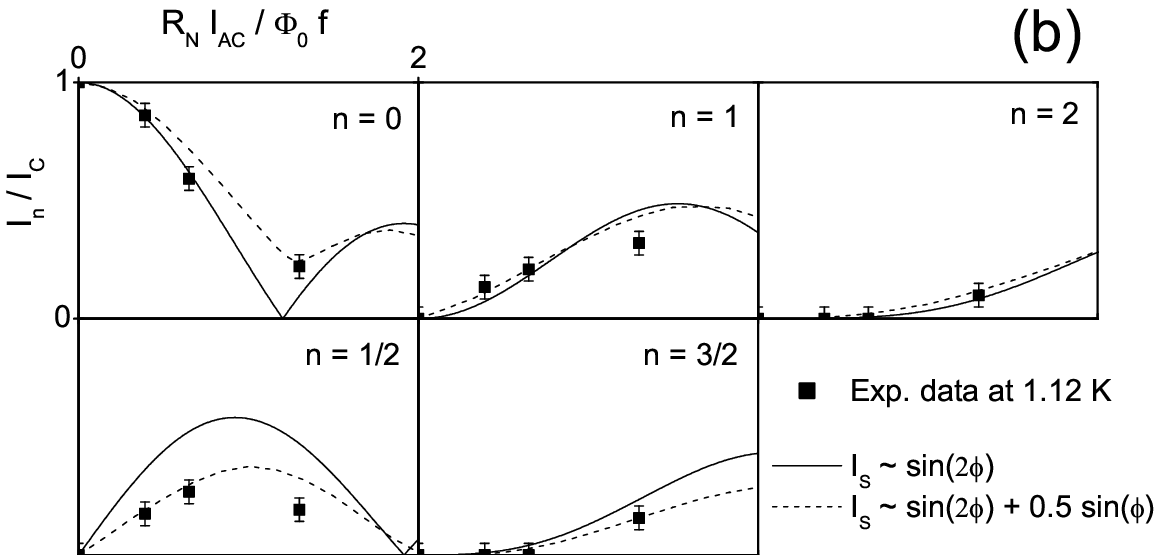}
\caption{(a) Shapiro steps in the voltage-current curves of the 17
nm thick junction at 1.12 K with an excitation at 800 kHz.
Excitation amplitudes are about 18, 9, 5 and 0 $\mu$A. The
respective curves are shifted by 0, 5, 10 and 15 $\mu$A for
clarity. (b) Width of the integer and half-integer steps {\it
versus} normalized amplitude of the alternative current.
Experimental values (symbols) are compared with numerical
simulations for two current-phase relations containing a dominant
$\sin(2\phi)$ component (lines). All graphs have same axis and
scales.} \label{fig:steps}
\end{figure}

We now investigate in more detail the harmonic composition of the
supercurrent in the 17 nm thick junction at 1.12 K. For this
purpose we measure the voltage-current curve for different
excitation amplitudes (Fig.~\ref{fig:steps}a) and compare
quantitatively the width of the Shapiro steps to the prediction of
the RSJ model for non-$\sin(\phi)$ current-phase relations
(Fig.~\ref{fig:steps}b). During the experiment we measured only
the relative amplitudes of the alternative current, but for a
quantitative comparison we need the absolute values $I_{AC}$.
Since no half-integer step is observed at 1.07 K we can assume a
$\sin(\phi)$ relation and adjust the step widths to the result of
the RSJ model, yielding an excitation amplitude of 18 $\mu$A. In
this case the normalized half-width $I_n/I_C$ (the full width is
$2I_n$) is equal to the Bessel function $J_n(a)$ when
$R_NI_C<\Phi_0f$ and where $a=R_NI_{AC}/\Phi_0f$
\cite{likharev-79-rmp}. Numerical simulations of the
voltage-current curves have been performed using a $\sin(2\phi)$
current-phase relation and the dependence of the step widths with
the excitation amplitude has been extracted. As expected the steps
are proportional to the Bessel functions $J_{2n}(2a)$. In
particular the n=1/2 step has the same behavior as the n=1 step of
a usual junction with an excitation two times larger. This n=1/2
step is not a sub-harmonic step, but the fundamental step for a
double Josephson frequency. Experimentally the width of this n=1/2
step first increases and then decreases with the excitation
amplitude in agreement with the oscillating behavior of the Bessel
functions. However its width is significantly smaller than
expected for a pure $\sin(2\phi)$ current-phase relation
(Fig.~\ref{fig:steps}b, solid line). This difference may be the
consequence of a residual $\sin(\phi)$ component in the
supercurrent. Numerical simulations have been performed to
calculate the step widths for current-phase relations containing
the two phase dependencies with different ratios. The best fit to
the experimental data is obtained when the amplitude of the
$\sin(\phi)$ is half that of the $\sin(2\phi)$ equal to 3.2 $\mu$A
(dashed line). This analysis indicates that the crossover
temperature, where the $\sin(\phi)$ component should disappear
completely, is not exactly 1.12 K. From the slope of the critical
current {\it versus} temperature (Fig.~\ref{fig:ict}) and the
amplitude of the residual $\sin(\phi)$ component (1.6 $\mu$A) we
can however estimate the actual crossover to be only 12 mK above
or below this temperature.

In summary we studied the finite voltage behavior of S/F/S
junctions under high frequency excitation and observed
half-integer Shapiro steps at the temperature of a 0\,--\,$\pi$
crossover where the critical current is non-zero. These steps
reveal the $\sin(2\phi)$ dependence of the current-phase relation
which is explained by the specific level splitting realized at the
crossover. This unusual relation changes rapidly for a
$\sin(\phi)$ dependence when one moves away from the crossover
temperature.

We thank M. Aprili and D. Est\`eve for pointing out the relevance
of the Shapiro steps to analyze the residual critical current. We
thank T.M. Klapwijk for the detailed analysis of the manuscript.
H. Sellier acknowledges the support of the Dutch foundation for
Fundamental Research on Matter (FOM).


\begin{thebibliography}{18}

\expandafter\ifx\csname
natexlab\endcsname\relax\def\natexlab#1{#1}\fi
\expandafter\ifx\csname bibnamefont\endcsname\relax
  \def\bibnamefont#1{#1}\fi
\expandafter\ifx\csname bibfnamefont\endcsname\relax
  \def\bibfnamefont#1{#1}\fi
\expandafter\ifx\csname citenamefont\endcsname\relax
  \def\citenamefont#1{#1}\fi
\expandafter\ifx\csname url\endcsname\relax
  \def\url#1{\texttt{#1}}\fi
\expandafter\ifx\csname
urlprefix\endcsname\relax\def\urlprefix{URL }\fi
\providecommand{\bibinfo}[2]{#2}
\providecommand{\eprint}[2][]{\url{#2}}

\bibitem[{\citenamefont{Buzdin et~al.}(1982)\citenamefont{Buzdin,
  Bulaevski\u{\i}, and Panyukov}}]{buzdin-82-jl}
\bibinfo{author}{\bibfnamefont{A.~I.} \bibnamefont{Buzdin}},
  \bibinfo{author}{\bibfnamefont{L.~N.} \bibnamefont{Bulaevski\u{\i}}},
  \bibnamefont{and} \bibinfo{author}{\bibfnamefont{S.~V.}
  \bibnamefont{Panyukov}}, \bibinfo{journal}{Pis'ma Zh. Eksp. Teor. Fiz.}
  \textbf{\bibinfo{volume}{35}}, \bibinfo{pages}{147} (\bibinfo{year}{1982}),
  \bibinfo{note}{[JETP Lett. {\bf 35}, 178 (1982)]}.

\bibitem[{\citenamefont{Dobrosavljevi\'c-Gruji\'c
  et~al.}(2000)\citenamefont{Dobrosavljevi\'c-Gruji\'c, Ziki\'c, and
  Radovi\'c}}]{dobrosavljevic-00-pc}
\bibinfo{author}{\bibfnamefont{L.}~\bibnamefont{Dobrosavljevi\'c-Gruji\'c}},
  \bibinfo{author}{\bibfnamefont{R.}~\bibnamefont{Ziki\'c}}, \bibnamefont{and}
  \bibinfo{author}{\bibfnamefont{Z.}~\bibnamefont{Radovi\'c}},
  \bibinfo{journal}{Physica C} \textbf{\bibinfo{volume}{331}},
  \bibinfo{pages}{254} (\bibinfo{year}{2000}).

\bibitem[{\citenamefont{Golubov et~al.}(2002)\citenamefont{Golubov, Kupriyanov,
  and Fominov}}]{golubov-02-jl}
\bibinfo{author}{\bibfnamefont{A.~A.} \bibnamefont{Golubov}},
  \bibinfo{author}{\bibfnamefont{M.~Y.} \bibnamefont{Kupriyanov}},
  \bibnamefont{and} \bibinfo{author}{\bibfnamefont{Y.~V.}
  \bibnamefont{Fominov}}, \bibinfo{journal}{Pis'ma Zh. Eksp. Teor. Fiz.}
  \textbf{\bibinfo{volume}{75}}, \bibinfo{pages}{709} (\bibinfo{year}{2002}),
  \bibinfo{note}{[JETP Lett. {\bf 75}, 588 (2002)]}.

\bibitem[{\citenamefont{Sellier et~al.}(2003)\citenamefont{Sellier, Baraduc,
  Lefloch, and Calemczuk}}]{sellier-03-prb}
\bibinfo{author}{\bibfnamefont{H.}~\bibnamefont{Sellier}},
  \bibinfo{author}{\bibfnamefont{C.}~\bibnamefont{Baraduc}},
  \bibinfo{author}{\bibfnamefont{F.}~\bibnamefont{Lefloch}}, \bibnamefont{and}
  \bibinfo{author}{\bibfnamefont{R.}~\bibnamefont{Calemczuk}},
  \bibinfo{journal}{Phys. Rev. B} \textbf{\bibinfo{volume}{68}},
  \bibinfo{pages}{054531} (\bibinfo{year}{2003}).

\bibitem[{\citenamefont{Radovi\'c et~al.}(2001)\citenamefont{Radovi\'c,
  Dobrosavljevi\'c-Gruji\'c, and Vuji\v{c}i\'c}}]{radovic-01-prb}
\bibinfo{author}{\bibfnamefont{Z.}~\bibnamefont{Radovi\'c}},
  \bibinfo{author}{\bibfnamefont{L.}~\bibnamefont{Dobrosavljevi\'c-Gruji\'c}},
  \bibnamefont{and}
  \bibinfo{author}{\bibfnamefont{B.}~\bibnamefont{Vuji\v{c}i\'c}},
  \bibinfo{journal}{Phys. Rev. B} \textbf{\bibinfo{volume}{63}},
  \bibinfo{pages}{214512} (\bibinfo{year}{2001}).

\bibitem[{\citenamefont{Heikkil{\"a} et~al.}(2000)\citenamefont{Heikkil{\"a},
  Wilhelm, and Sch{\"o}n}}]{heikkila-00-el}
\bibinfo{author}{\bibfnamefont{T.~T.} \bibnamefont{Heikkil{\"a}}},
  \bibinfo{author}{\bibfnamefont{F.~K.} \bibnamefont{Wilhelm}},
  \bibnamefont{and}
  \bibinfo{author}{\bibfnamefont{G.}~\bibnamefont{Sch{\"o}n}},
  \bibinfo{journal}{Europhys. Lett.} \textbf{\bibinfo{volume}{51}},
  \bibinfo{pages}{434} (\bibinfo{year}{2000}).

\bibitem[{\citenamefont{Ryazanov
  et~al.}(2001{\natexlab{a}})\citenamefont{Ryazanov, Oboznov, Rusanov,
  Veretennikov, Golubov, and Aarts}}]{ryazanov-01-prl}
\bibinfo{author}{\bibfnamefont{V.~V.} \bibnamefont{Ryazanov}},
  \bibinfo{author}{\bibfnamefont{V.~A.} \bibnamefont{Oboznov}},
  \bibinfo{author}{\bibfnamefont{A.~Y.} \bibnamefont{Rusanov}},
  \bibinfo{author}{\bibfnamefont{A.~V.} \bibnamefont{Veretennikov}},
  \bibinfo{author}{\bibfnamefont{A.~A.} \bibnamefont{Golubov}},
  \bibnamefont{and} \bibinfo{author}{\bibfnamefont{J.}~\bibnamefont{Aarts}},
  \bibinfo{journal}{Phys. Rev. Lett.} \textbf{\bibinfo{volume}{86}},
  \bibinfo{pages}{2427} (\bibinfo{year}{2001}{\natexlab{a}}).

\bibitem[{\citenamefont{Kontos et~al.}(2002)\citenamefont{Kontos, Aprili,
  Lesueur, Gen\^et, Stephanidis, and Boursier}}]{kontos-02-prl}
\bibinfo{author}{\bibfnamefont{T.}~\bibnamefont{Kontos}},
  \bibinfo{author}{\bibfnamefont{M.}~\bibnamefont{Aprili}},
  \bibinfo{author}{\bibfnamefont{J.}~\bibnamefont{Lesueur}},
  \bibinfo{author}{\bibfnamefont{F.}~\bibnamefont{Gen\^et}},
  \bibinfo{author}{\bibfnamefont{B.}~\bibnamefont{Stephanidis}},
  \bibnamefont{and} \bibinfo{author}{\bibfnamefont{R.}~\bibnamefont{Boursier}},
  \bibinfo{journal}{Phys. Rev. Lett.} \textbf{\bibinfo{volume}{89}},
  \bibinfo{pages}{137007} (\bibinfo{year}{2002}).

\bibitem[{\citenamefont{Ryazanov
  et~al.}(2001{\natexlab{b}})\citenamefont{Ryazanov, Oboznov, Veretennikov, and
  Rusanov}}]{ryazanov-01-prb}
\bibinfo{author}{\bibfnamefont{V.~V.} \bibnamefont{Ryazanov}},
  \bibinfo{author}{\bibfnamefont{V.~A.} \bibnamefont{Oboznov}},
  \bibinfo{author}{\bibfnamefont{A.~V.} \bibnamefont{Veretennikov}},
  \bibnamefont{and} \bibinfo{author}{\bibfnamefont{A.~Y.}
  \bibnamefont{Rusanov}}, \bibinfo{journal}{Phys. Rev. B}
  \textbf{\bibinfo{volume}{65}}, \bibinfo{pages}{020501}
  (\bibinfo{year}{2001}{\natexlab{b}}).

\bibitem[{\citenamefont{Guichard et~al.}(2003)\citenamefont{Guichard, Aprili,
  Bourgeois, Kontos, Lesueur, and Gandit}}]{guichard-03-prl}
\bibinfo{author}{\bibfnamefont{W.}~\bibnamefont{Guichard}},
  \bibinfo{author}{\bibfnamefont{M.}~\bibnamefont{Aprili}},
  \bibinfo{author}{\bibfnamefont{O.}~\bibnamefont{Bourgeois}},
  \bibinfo{author}{\bibfnamefont{T.}~\bibnamefont{Kontos}},
  \bibinfo{author}{\bibfnamefont{J.}~\bibnamefont{Lesueur}}, \bibnamefont{and}
  \bibinfo{author}{\bibfnamefont{P.}~\bibnamefont{Gandit}},
  \bibinfo{journal}{Phys. Rev. Lett.} \textbf{\bibinfo{volume}{90}},
  \bibinfo{pages}{167001} (\bibinfo{year}{2003}).

\bibitem[{\citenamefont{Hamilton and Johnson}(1972)}]{hamilton-72-pla}
\bibinfo{author}{\bibfnamefont{C.~A.} \bibnamefont{Hamilton}} \bibnamefont{and}
  \bibinfo{author}{\bibfnamefont{E.~G.} \bibnamefont{Johnson}},
  \bibinfo{journal}{Phys. Lett. A} \textbf{\bibinfo{volume}{41}},
  \bibinfo{pages}{393} (\bibinfo{year}{1972}).

\bibitem[{\citenamefont{Bardeen and Johnson}(1972)}]{bardeen-72-prb}
\bibinfo{author}{\bibfnamefont{J.}~\bibnamefont{Bardeen}} \bibnamefont{and}
  \bibinfo{author}{\bibfnamefont{J.~L.} \bibnamefont{Johnson}},
  \bibinfo{journal}{Phys. Rev. B} \textbf{\bibinfo{volume}{5}},
  \bibinfo{pages}{72} (\bibinfo{year}{1972}).

\bibitem[{\citenamefont{Kulik and Omelyanchuk}(1975)}]{kulik-75-jl}
\bibinfo{author}{\bibfnamefont{I.~O.} \bibnamefont{Kulik}} \bibnamefont{and}
  \bibinfo{author}{\bibfnamefont{A.~N.} \bibnamefont{Omelyanchuk}},
  \bibinfo{journal}{Zh. Eksp. Teor. Fiz. Pis'ma Red.}
  \textbf{\bibinfo{volume}{21}}, \bibinfo{pages}{216} (\bibinfo{year}{1975}),
  \bibinfo{note}{[JETP Lett. {\bf 21}, 96 (1975)]}.

\bibitem[{\citenamefont{Wilhelm et~al.}(1998)\citenamefont{Wilhelm, Sch{\"o}n,
  and Zaikin}}]{wilhelm-98-prl}
\bibinfo{author}{\bibfnamefont{F.~K.} \bibnamefont{Wilhelm}},
  \bibinfo{author}{\bibfnamefont{G.}~\bibnamefont{Sch{\"o}n}},
  \bibnamefont{and} \bibinfo{author}{\bibfnamefont{A.~D.}
  \bibnamefont{Zaikin}}, \bibinfo{journal}{Phys. Rev. Lett.}
  \textbf{\bibinfo{volume}{81}}, \bibinfo{pages}{1682} (\bibinfo{year}{1998}).

\bibitem[{\citenamefont{Clarke}(1968)}]{clarke-68-prl}
\bibinfo{author}{\bibfnamefont{J.}~\bibnamefont{Clarke}},
  \bibinfo{journal}{Phys. Rev. Lett.} \textbf{\bibinfo{volume}{21}},
  \bibinfo{pages}{1566} (\bibinfo{year}{1968}).

\bibitem[{\citenamefont{Lehnert et~al.}(1999)\citenamefont{Lehnert, Argaman,
  Blank, Wong, Allen, Hu, and Kroemer}}]{lehnert-99-prl}
\bibinfo{author}{\bibfnamefont{K.~W.} \bibnamefont{Lehnert}},
  \bibinfo{author}{\bibfnamefont{N.}~\bibnamefont{Argaman}},
  \bibinfo{author}{\bibfnamefont{H.-R.} \bibnamefont{Blank}},
  \bibinfo{author}{\bibfnamefont{K.~C.} \bibnamefont{Wong}},
  \bibinfo{author}{\bibfnamefont{S.~J.} \bibnamefont{Allen}},
  \bibinfo{author}{\bibfnamefont{E.~L.} \bibnamefont{Hu}}, \bibnamefont{and}
  \bibinfo{author}{\bibfnamefont{H.}~\bibnamefont{Kroemer}},
  \bibinfo{journal}{Phys. Rev. Lett.} \textbf{\bibinfo{volume}{82}},
  \bibinfo{pages}{1265} (\bibinfo{year}{1999}).

\bibitem[{\citenamefont{Dubos et~al.}(2001)\citenamefont{Dubos, Courtois,
  Buisson, and Pannetier}}]{dubos-01-prl}
\bibinfo{author}{\bibfnamefont{P.}~\bibnamefont{Dubos}},
  \bibinfo{author}{\bibfnamefont{H.}~\bibnamefont{Courtois}},
  \bibinfo{author}{\bibfnamefont{O.}~\bibnamefont{Buisson}}, \bibnamefont{and}
  \bibinfo{author}{\bibfnamefont{B.}~\bibnamefont{Pannetier}},
  \bibinfo{journal}{Phys. Rev. Lett.} \textbf{\bibinfo{volume}{87}},
  \bibinfo{pages}{206801} (\bibinfo{year}{2001}).

\bibitem[{\citenamefont{Likharev}(1979)}]{likharev-79-rmp}
\bibinfo{author}{\bibfnamefont{K.~K.} \bibnamefont{Likharev}},
  \bibinfo{journal}{Rev. Mod. Phys.} \textbf{\bibinfo{volume}{51}},
  \bibinfo{pages}{101} (\bibinfo{year}{1979}).

\end{thebibliography}

\end{document}